\documentclass[12pt]{article}
\advance\textwidth by +1.0in
\advance\textheight by +1.0in
\advance\oddsidemargin by -0.5in
\advance\evensidemargin by -1.0in
\advance\topmargin by -0.5in
\def\wt#1{\widetilde{#1}}
\def\vb#1{\mbox{\boldmath$#1$}}
\def\pd#1#2{\frac{\partial #1}{\partial #2}}
\def\wh#1{\widehat{#1}}
\def\bdot{\,\vb{\cdot}\,}
\def\btimes{\,\vb{\times}\,}
\def\bhat{\wh{{\sf b}}}

\newcommand{\bc}{\begin{center}}
\newcommand{\ec}{\end{center}}

\newcommand{\be}{\begin{eqnarray*}}
\newcommand{\ee}{\end{eqnarray*}}
\newcommand{\bt}{\begin{tabbing}}
\newcommand{\et}{\end{tabbing}}
\newcommand{\no}{\noindent}

\begin{document}

\begin{center}

{\large {\bf A guiding-center Fokker-Planck collision operator}} \\
\vspace*{0.1in}
{\large {\bf for nonuniform magnetic fields}} \\

\vspace*{0.4in}

Alain J.~Brizard \\
\vspace*{0.2in}
{\it Department of Chemistry and Physics, Saint Michael's College} \\
{\it Colchester, Vermont 05439}

\end{center}

A new formulation for collisional kinetic theory is presented based on the use of Lie-transform methods to eliminate fast orbital time scales from a general bilinear collision operator. As an application of this new formalism, a general guiding-center bilinear Fokker-Planck (FP) collision operator is derived following the elimination of the fast gyromotion time scale of a charged particle moving in a nonuniform magnetic field. It is expected that classical transport processes in a strongly magnetized nonuniform plasma can, thus, be described in terms of this reduced guiding-center FP kinetic theory. The present paper introduces the reduced-collision formalism only, while its applications are left to future work.

\vspace*{0.2in}

\no

\vfill\eject

\no
{\sf I. INTRODUCTION}

\vspace*{0.2in}

Lie-transform perturbation methods are normally associated with the asymptotic elimination of fast time scales from single-particle Hamilton equations \cite{Littlejohn_82,Brizard_01}. In the present paper, the general Lie-transform rules for the transformation of an arbitrary bilinear collision operator are presented. 

As a first step, Lie-transform methods are used to asymptotically eliminate a fast orbital time scale from the collisionless (Vlasov) evolution operator of a collisional kinetic equation. This dynamical reduction is performed through a phase-space coordinate transformation, from which collisionless reduced Vlasov kinetic theories are derived (e.g., gyrokinetic Vlasov theory \cite{Brizard_89,Brizard_00}). As a result of this coordinate transformation, the collision operator is transformed into a new collision operator in which the fast time-scale dependence is still present; this feature is common to all collisional theories based on the use reduced guiding-center coordinates \cite{Balescu,CT,Auerbach}. To eliminate this residual fast-time-scale dependence, we proceed with a second step involving a fast-angle averaging procedure and a closure scheme based on a time-scale ordering involving the characteristic collisional time scale. A {\it reduced} collisional kinetic equation is, therefore, obtained in which the fast-angle-averaged distribution evolves in a reduced phase space. Note that fast-angle averaging is required only in the presence of collisions since the first step involves a {\it reversible} phase-space coordinate transformation that requires no averaging.

The present approach is to be contrasted with the traditional multiple time-scale approach \cite{MTS}, whereby the collisional kinetic equation in particle phase space is expanded in powers of a small parameter $\epsilon$, with the particle distribution function expanded as $f = \sum_{n = 0}\,\epsilon^{n}\,f_{n}$. In general, the zeroth-order component $f_{0}$ is independent of the fast time scale and the fast-time-scale dependence of $f_{n}$ involves contributions from collisionless (Hamiltonian) dynamics and collisional dynamics, which introduces several layers of tedious algebra before a reduced collisional kinetic equation is derived \cite{HTH,PHR}, with a {\it linearized} collision operator.

By using Lie-transform methods, we obtain simpler and more compact expressions for transformed guiding-center Fokker-Planck (FP) collision operators (for uniform and nonuniform magnetic fields) when compared to those obtained by the standard approach for uniform magnetic fields \cite{CT,Auerbach}. The present approach is expected to be appropriate appropriate for applications in collisional gyrokinetic theory and gyrokinetic particle simulations 
\cite{XR,DC}. 

\vspace*{0.2in}

\no
{\bf A. Brief Introduction to Collisional Plasma Kinetic Theory}

\vspace*{0.2in}

We begin with a brief introduction of collisional plasma kinetic theory (see Refs.~\cite{Braginskii}-\cite{Hinton}, for example). The study of irreversible transport processes in nonuniform plasmas is based on solutions of a collisional plasma kinetic equation
\begin{equation}
\frac{df}{dt}({\bf z},t) \;=\; {\cal C}[f]({\bf z},t),
\label{eq:cke}
\end{equation}
where $f$ is the {\it test}-particle distribution function and ${\bf z}$ are coordinates in the six-dimensional single-particle phase space. The Vlasov operator
\begin{equation}
\frac{d}{dt} \;\equiv\; \pd{}{t} \;+\; \dot{z}^{\alpha}({\bf z},t) \;\pd{}{z^{\alpha}}
\label{eq:vlas_op}
\end{equation}
characterizes the {\it dissipationless} time evolution of $f$, where the Hamiltonian particle orbit 
${\bf z}(t;{\bf z}_{0})$ is the solution of the Hamilton equations $\dot{z}^{\alpha}({\bf z},t) \equiv \{ z^{\alpha},\; 
h({\bf z},t)\}$ with the initial condition ${\bf z}_{0}$; here, $h$ and $\{\,,\,\}$ are the single-particle Hamiltonian and Poisson bracket, respectively. (Summation over repeated indices is implied throughout the paper; greek indices take values from $1$ to $6$, while latin indices take values from $1$ to $3$.) 

In the absence of collisions $({\cal C} \equiv 0)$, Eq.~(\ref{eq:cke}) is the Vlasov equation \cite{Vlasov}: 
$df({\bf z},t)/dt = 0$, whose solution $f({\bf z},t)$ is constant along a Hamiltonian particle orbit. The collision operator 
\begin{equation}
{\cal C}[f] \;\equiv\; \sum\; {\cal C}[f;f']
\label{eq:coll_sum}
\end{equation}
in Eq.~(\ref{eq:cke}) characterizes the {\it dissipative} (irreversible) time evolution of $f$, where the {\it bilinear} collision operator ${\cal C}[f;f']$ describes binary collisions between test-particles and {\it field}-particles (with distribution $f'$); the summation is over all field-particle species (and includes like-particle collisions).

Solutions of the collisional kinetic equation (\ref{eq:cke}) are quite difficult to obtain in general and various approximation schemes must be adopted in order to arrive at useful solutions. One such approximation scheme involves removing fast {\it orbital} time scales associated with Hamiltonian particle orbits in six-dimensional phase space \cite{Balescu,Bernstein_Molvig}. Here, the motion of magnetically-confined charged particles exhibits three distinct orbital time scales $(\tau_{g} \ll \tau_{b} \ll \tau_{d})$: (1) a fast {\it gyration} time scale $\tau_{g}$ associated with the gyromotion of charged particles about a magnetic field line; (2) an intermediate {\it bounce} time scale $\tau_{b}$ associated with the periodic parallel motion of charged particle along a magnetic field line; and (3) a slow {\it drift} time scale $\tau_{d}$ associated with the perpendicular motion of charged particles across {\it nonuniform} magnetic field lines. 

After choosing a {\it slow} time scale of interest (labeled $\tau$), we identify all {\it fast} time scales $\tau_{k}$ which satisfy the condition 
$\tau_{k} \ll \tau$. For example, if we are interested in plasma dynamics on the bounce time scale $(\tau \sim \tau_{b})$ then the gyromotion time scale is considered fast $(\tau_{g} \ll \tau)$ and it can be asymptotically removed. For each such fast orbital time scale $\tau_{k}$, a pair of action-angle variables $(J_{k},\theta^{k})$ is assigned: $\theta^{k}$ is the fast angle variable (with $\dot{\theta}^{k} \equiv \Omega^{k}$ denoting the fast orbital frequency) and its canonically-conjugate action $J_{k}$ is an adiabatic invariant on a time scale longer than $\tau_{k} \equiv \oint 
d\theta^{k}/\Omega^{k}$ (i.e., the fast-angle average of $\dot{J}_{k}$ is zero). The asymptotic elimination of a fast orbital time scale $\tau_{k}$ from the Vlasov operator (2) can be carried out to arbitrary order in the small parameter $\epsilon_{k} \equiv \omega/\Omega^{k} \ll 1$ (where $i\,\partial_{\tau} \rightarrow \omega$ denotes a characteristic frequency) \cite{Littlejohn_82,Kruskal}. 

In the present paper, we focus our attention on the asymptotic elimination of the gyromotion time scale leading to the reduced guiding-center dynamics of charged particles in nonuniform magnetic fields. Here, the reduced guiding-center coordinates $({\bf X}, {\cal E}, \mu, \theta)$ include the guiding-center position ${\bf X}$, the guiding-center kinetic energy ${\cal E}$, the guiding-center magnetic moment $\mu$, and the guiding-center gyro-angle $\theta$. Each coordinate is expressed as an asymptotic expansion in powers of the dimensionless parameter $\epsilon_{B} = \rho/L_{B}$ 
\cite{Littlejohn_79,gc,Brizard_95}, where $\rho$ is the characteristic gyroradius of a charged particle and $L_{B}$ denotes the magnetic field nonuniformity length scale.

\vspace*{0.2in}

\no
{\bf B. Lie-transform Phase-space Transformations}

\vspace*{0.2in}

The standard mathetical method used to asymptotically eliminate fast time scales is the Lie-transform method, which we briefly review here. In particular, we show the transformation rules by which each element of the collisional kinetic equation (\ref{eq:cke}) (i.e., $f$, $d/dt$, and ${\cal C}$) transform under a near-identity transformation. 

Using Lie-transform methods \cite{Littlejohn_82}, the process by which a fast time scale is removed from 
Eq.~(\ref{eq:vlas_op}) involves a near-identity transformation on particle phase space: 
\begin{equation}
{\cal T}_{\epsilon}:\; {\bf z} \;\rightarrow\; {\bf Z}({\bf z};\epsilon) \;\equiv\; {\cal T}_{\epsilon}{\bf z},\;\; {\rm with}\;\; {\bf Z}({\bf z};0) \;=\; {\bf z},
\label{eq:near_trans}
\end{equation}
where $\epsilon \ll 1$ is a small parameter (for reduced guiding-center dynamics, $\epsilon = \epsilon_{B}$). In what follows, the new six-dimensional phase-space coordinates ${\bf Z} \equiv ({\bf Z}_{{\rm R}};J,\theta)$ are divided into the {\it reduced} phase-space coordinates ${\bf Z}_{{\rm R}}$ (the {\it reduced} particle dynamics takes place in a {\it four}-dimensional space) and the fast canonical action-angle pair $(J,\theta)$. 

Next, using the transformation (\ref{eq:near_trans}), we define the {\it push-forward} operator on scalar fields 
\cite{Littlejohn_82} {\it induced} by the near-identity transformation (\ref{eq:near_trans}):
\begin{equation}
{\sf T}_{\epsilon}^{-1}:\; f \;\rightarrow\; F \;\equiv\; {\sf T}_{\epsilon}^{-1}f,
\label{eq:pushforward}
\end{equation}
i.e., ${\sf T}_{\epsilon}^{-1}$ transforms a scalar field $f$ on the phase space with coordinates ${\bf z}$ into a scalar field $F$ on the phase space with coordinates ${\bf Z}$: $F({\bf Z}) = {\sf T}_{\epsilon}^{-1}f({\bf Z}) = 
f({\cal T}_{\epsilon}^{-1}{\bf Z}) = f({\bf z})$. Since the transformation (\ref{eq:near_trans}) is invertible, i.e., there exists an inverse near-identity transformation 
\begin{equation}
{\cal T}_{\epsilon}^{-1}:\; {\bf Z} \;\rightarrow\; {\bf z}({\bf Z};\epsilon) \;\equiv\; {\cal T}_{\epsilon}^{-1}{\bf Z},\;\; {\rm with}\;\; {\bf z}({\bf Z};0) \;=\; 
{\bf Z},
\label{eq:near_inv}
\end{equation}
we also define the {\it pull-back} operator \cite{Littlejohn_82}: 
\begin{equation}
{\sf T}_{\epsilon}:\; F \;\rightarrow\; f \;\equiv\; {\sf T}_{\epsilon}F,
\label{eq:pullback}
\end{equation}
i.e., ${\sf T}_{\epsilon}$ transforms a scalar field $F$ on the phase space with coordinates ${\bf Z}$ into a scalar field $f$ on the phase space with coordinates ${\bf z}$: $f({\bf z}) = {\sf T}_{\epsilon}F({\bf z}) = 
F({\cal T}_{\epsilon}{\bf z}) = F({\bf Z})$. 

Lastly, we consider the transformation of a general operator ${\cal L}: f \rightarrow {\cal L}f$ acting {\it locally} on the particle distribution $f$ such that ${\cal L}f$ is also a scalar field on particle phase space. Hence, the induced transformation of the scalar field ${\cal L}f$ follows the push-forward rule (\ref{eq:pushforward}): ${\cal L}f \rightarrow {\sf T}_{\epsilon}^{-1}{\cal L}f$. By combining this push-forward transformation with the pull-back transformation $f = {\sf T}_{\epsilon}F$, we construct a new local operator ${\cal L}_{\epsilon}: F \rightarrow 
{\cal L}_{\epsilon}F$ defined as
\begin{equation}
{\cal L}_{\epsilon}F \;\equiv\; \left. \left. {\sf T}_{\epsilon}^{-1} {\cal L}\right( {\sf T}_{\epsilon}\;F \right).
\label{eq:pushpull_operator}
\end{equation}
This induced transformation rule for operators will be used in the next Section to transform the Vlasov operator $d/dt$ and the collision operator ${\cal C}$ appearing in the collisional kinetic equation (\ref{eq:cke}).

\vspace*{0.2in}

\no
{\bf C. Organization}

\vspace*{0.2in}

The remainder of the paper is organized as follows. In Sec.~II, we investigate the fast-time dependence of the transformed collisional kinetic equation and discover that, although the fast time scale has been eliminated from the dissipationless (Vlasov) part of the kinetic equation, the dissipative (collisional) part has retained fast-time dependence as a result of collisions. This is not too surprising after all since these phase-space transformations were never designed for that purpose. This means that the transformed collision operator is not yet {\it reduced} in the sense that the fast-angle-independent (averaged) part, denoted $\langle F\rangle$, and the fast-angle-dependent part, denoted $\wt{F} = F - \langle F\rangle$, of the transformed distribution function $F$ are still coupled by collisions. The reduction of the transformed collision operator involves a closure scheme yielding a reduced collision operator acting on $\langle F\rangle$ alone. This closure scheme is based on an asymptotic expansion based on a new expansion parameter $\epsilon_{\nu} = \nu/\Omega = \rho/\lambda_{\nu} \ll 1$ defined as the ratio of the characteristic collision frequency $\nu$ over the fast frequency 
$\Omega$ or the ratio of the gyroradius $\rho$ over the collisional mean-free-path $\lambda_{\nu}$. For most practical applications, however, we argue that only the lowest term in this $\epsilon_{\nu}$-expansion is relevant. 

The purpose of the present paper is, therefore, to derive first-order $(\epsilon_{B} = \rho/L_{B})$ corrections of a reduced guiding-center Fokker-Planck (FP) collision operator, which might be suitable for analytical and numerical study of collisional and turbulent transport processes in nonuniform magnetized plasmas. In Sec.~III, we derive the lowest-order term in a reduced Fokker-Planck (FP) collision operator. Starting from a standard FP collision operator (describing collisional drag and diffusion in momentum space), we derive an expression for a reduced FP collision operator in a reduced $(4 + 1)$ phase space with coordinates $({\bf Z}_{R},J)$. In particular, the reduced FP collision operator describes {\it spatial} drag and diffusion processes normally described from a collisional fluid-moment approach. 

As a specific example, we focus our attention in Sec.~IV on the guiding-center phase-space transformation associated with the asymptotic elimination of the fast gyro-motion time scale of a charged particle moving in a nonuniform magnetic field. General expressions are obtained for the collisional FP drag and diffusion coefficients in guiding-center phase space. For clarity of exposition, we also use an isotropic field-particle model introduced in Sec.~II to investigate how magnetic-field nonuniformity influences these guiding-center FP coefficients. We also discuss possible extensions of our work introduced by background nonuniform electric fields (which are outside the scope of the present work) and summarize our findings in Sec.~V. Lastly, we present some technical details about the guiding-center transformation for nonuniform magnetic fields in Appendix A.

\vspace*{0.2in}

\no
{\sf II. TRANSFORMED COLLISIONAL KINETIC EQUATION}

\vspace*{0.2in}

The transformation of the collisional kinetic equation (\ref{eq:cke}), induced by the near-identity phase-space transformation (\ref{eq:near_trans}), yields a new collisional kinetic equation:
\begin{equation}
\frac{d_{\epsilon}}{dt}\,F({\bf Z},t) \;=\; {\cal C}_{\epsilon}[F]({\bf Z},t) \;=\; \sum^{\prime}\; 
{\cal C}_{(\epsilon,\epsilon')}[F;F']({\bf Z},t),
\label{eq:rcke}
\end{equation}
where $F \equiv {\sf T}_{\epsilon}^{-1}f$ is the transformed test-particle distribution function obtained from the push-forward of the test-particle distribution function $f$, the transformed Vlasov operator $d_{\epsilon}/dt$ is defined from Eq.~(\ref{eq:pushpull_operator}) as
\begin{equation}
\frac{d_{\epsilon}}{dt}\,F({\bf Z},t) \;\equiv\; {\sf T}_{\epsilon}^{-1}\left( \frac{d}{dt} \,{\sf T}_{\epsilon}F \right)(
{\bf Z},t),
\label{eq:rvlas_op}
\end{equation}
while the transformed bilinear collision operator is defined as
\begin{equation}
{\cal C}_{(\epsilon,\epsilon')}[F;F']({\bf Z},t) \;\equiv\; \left. \left. {\sf T}_{\epsilon}^{-1} \right( {\cal C}[
{\sf T}_{\epsilon}F;\, {\sf T}_{\epsilon'}F' ]\right)({\bf Z},t),
\label{eq:tcoll_bi}
\end{equation}
where $\epsilon' \equiv \omega/\Omega'$ is the small parameter associated with the field-particle species. We note that the bilinearity of the collision operator (\ref{eq:coll_sum}) allows a separate treatment of the test-particle ($f$) and field-particle ($f^{\prime}$) distributions and, for convenience of notation, we, henceforth, write the transformed collision operator (\ref{eq:tcoll_bi}) as
\begin{equation}
{\cal C}_{\epsilon}[F]({\bf Z},t) \;\equiv\; {\sf T}_{\epsilon}^{-1}\left( {\cal C}[{\sf T}_{\epsilon}F] 
\right)({\bf Z},t),
\label{eq:rcoll_op}
\end{equation}
and omit mention of the field-particle species unless otherwise needed.

The transformed Vlasov operator (\ref{eq:rvlas_op}) is now explicitly written in terms of the new phase-space coordinates 
$({\bf Z}_{R};\, J,\theta)$ as
\begin{equation} 
\frac{d_{\epsilon}}{dt} \;=\; \pd{}{t} \;+\; \dot{Z}_{\epsilon}^{\alpha}\,\pd{}{Z^{\alpha}} \;=\; \frac{d_{{\rm R}}}{dt} \;+\; \dot{\theta}\;\pd{}{\theta},
\label{eq:RHamOp}
\end{equation}
where the transformed Hamilton equations $\dot{Z}_{\epsilon}^{\alpha} \equiv \{ Z^{\alpha},\, H \}_{\epsilon}$ are defined in terms of the transformed Hamiltonian $H \equiv {\sf T}_{\epsilon}^{-1}h$ and the transformed Poisson bracket
\begin{equation}
\{ F,\;G \}_{\epsilon}({\bf Z}) \;\equiv\; \left. \left. {\sf T}_{\epsilon}^{-1}\right( \left\{ {\sf T}_{\epsilon}F,\; 
{\sf T}_{\epsilon}G \right\} \right)({\bf Z}).
\label{eq:r_PB}
\end{equation}
We note that the transformed Hamiltonian $H({\bf Z}_{R};\, J)$ and the transformed Poisson bracket $\{\,,\,\}_{\epsilon}$ are (by construction) independent of the fast angle $\theta$, so that Hamilton's equations $\dot{{\bf Z}}_{{\rm R}}({\bf Z}_{{\rm R}};J)$ and $\dot{\theta} = 
\Omega({\bf Z}_{{\rm R}};J)$ -- and, thus, the transformed Vlasov operator $d_{\epsilon}/dt$ -- are independent of the fast angle $\theta$ while $\dot{J} \equiv 0$ (to all orders in $\epsilon$).

The $\theta$-independence of the transformed Vlasov operator (\ref{eq:RHamOp}) allows the separation of the transformed collisional kinetic equation (\ref{eq:rcke}) into two coupled kinetic equations: one for the {\it $\theta$-averaged} distribution, denoted $\langle F\rangle$, and one for the {\it $\theta$-dependent} distribution, denoted $\wt{F} \equiv 
F - \langle F\rangle$. The collisional kinetic equation for $\langle F\rangle$ is obtained by $\theta$-averaging both sides of Eq.~(\ref{eq:rcke}):
\begin{equation}
\frac{d_{{\rm R}}}{dt}\; \langle F\rangle \;=\; \left\langle{\cal C}_{\epsilon}[F] \right\rangle \;\equiv\; \left\langle 
{\cal C}_{\epsilon}[\langle F\rangle]\right\rangle \;+\; \left\langle {\cal C}_{\epsilon}[\wt{F}]\right\rangle,
\label{eq:rcke_ind}
\end{equation}
where we used the $\theta$-independence of the reduced Vlasov operator $d_{{\rm R}}/dt$ on the left side of 
Eq.~(\ref{eq:rcke_ind}), while the collisional kinetic equation for $\wt{F}$ is obtained by subtracting 
Eq.~(\ref{eq:rcke_ind}) from Eq.~(\ref{eq:rcke}):
\begin{equation}
\left( \Omega\,\pd{}{\theta} \;+\; \frac{d_{{\rm R}}}{dt} \right)\wt{F} \;=\; {\cal C}_{\epsilon}[ F] \;-\; \left\langle 
{\cal C}_{\epsilon}[F] \right\rangle.
\label{eq:rcke_dep}
\end{equation}
Note that the term $\langle{\cal C}_{\epsilon}[\wt{F}]\rangle$ in Eq.~(\ref{eq:rcke_ind}) does not vanish in general and, thus, the collisional kinetic equations (\ref{eq:rcke_ind}) and (\ref{eq:rcke_dep}) are coupled through collisions (i.e., when ${\cal C} = 0$, it can be shown that $\wt{F} = 0$ to all orders in $\epsilon$). Hence, Eq.~(\ref{eq:rcke_ind}) still contains a dependence on the fast angle $\theta$ through the $\theta$-dependent part $\wt{F}$ of the transformed distribution $F$.

The ultimate goal in reduced collisional kinetic theory is to transform Eq.~(\ref{eq:rcke_ind}) into a {\it closed} collisional kinetic equation:
\begin{equation}
\frac{d_{{\rm R}}}{dt}\langle F\rangle({\bf Z}_{{\rm R}},t;J) \;\equiv\; {\cal C}_{{\rm R}}[\langle F\rangle]
({\bf Z}_{{\rm R}},t;J),
\label{eq:rcke_closed}
\end{equation}
where the reduced (R) collision operator ${\cal C}_{{\rm R}}$ acts on the $\theta$-independent part $\langle F\rangle$ alone. To obtain this collision operator, we must solve the collisional kinetic equation (\ref{eq:rcke_dep}) for $\wt{F}$ 
in terms of $\langle F \rangle$ and substitute this functional solution into $\langle{\cal C}_{\epsilon}[\wt{F}]\rangle$. This is accomplished by expanding the transformed distribution function $F$ as a Fourier series in the fast orbital angle $\theta$: 
\begin{equation}
F \;\equiv\; \langle F\rangle \;+\; \sum_{\ell > 0}\, \left( \wt{F}_{\ell}\,e^{i\ell\theta} \;+\; \wt{F}_{\ell}^{*}\,
e^{-i\ell\theta} \right),
\label{eq:Fourier}
\end{equation}
where $\wt{F}_{\ell}^{*} \equiv \wt{F}_{-\ell}$, and defining the averaged collision operators
\begin{equation}
{\cal C}_{\ell}^{k}[g] \;\equiv\; \left\langle e^{-i\ell\theta}\,{\cal C}_{\epsilon}\left[g\,e^{ik\theta}\right] \right\rangle, 
\label{eq:Clk_def}
\end{equation}
where $g$ is an arbitrary $\theta$-independent function and $(k,\ell)$ are arbitrary integers. By substituting these expressions into Eqs.~(\ref{eq:rcke_ind})-(\ref{eq:rcke_dep}), respectively, we obtain
\begin{eqnarray}
\frac{d_{{\rm R}}}{dt}\langle F\rangle & = & {\cal C}_{0}^{0}[\langle F\rangle] \;+\; \sum_{\ell\neq 0}\, 
{\cal C}_{0}^{\ell}[\wt{F}_{\ell}], 
\label{eq:rcke_0} \\
\left( i\ell\,\Omega + \frac{d_{{\rm R}}}{dt}\right) \wt{F}_{\ell} & = & {\cal C}_{\ell}^{0}[\langle F\rangle] \;+\; 
\sum_{k\neq 0}\, {\cal C}_{\ell}^{k}[\wt{F}_{k}].
\label{eq:rcke_l}
\end{eqnarray}
The approximate solution of Eq.~(\ref{eq:rcke_l}) for each integer $\ell \neq 0$ is based on the small parameter 
$\epsilon_{\nu} \equiv \nu/\Omega \ll 1$ (which is different from $\epsilon \equiv \omega/\Omega$), where $\nu$ is the characteristic collisional dissipation rate. To first order in $\epsilon_{\nu}$, and assuming that ${\cal C}_{\ell}^{0}
[\langle F\rangle] \neq 0$ in Eq.~(\ref{eq:rcke_l}), we, thus, find the functional solution $\wt{F}_{\ell}[\langle F \rangle] = (i\ell\Omega)^{-1}\;C_{\ell}^{0}[\langle F \rangle]$, which when substituted into the second term on the right side of Eq.~(\ref{eq:rcke_0}) yields the first two terms in the asymptotic expansion for 
${\cal C}_{{\rm R}}$:
\begin{equation}
{\cal C}_{{\rm R}}[\langle F\rangle] \;=\; {\cal C}_{0}^{0}[\langle F\rangle] \;-\; i\,\epsilon_{\nu}\;
\sum_{\ell\neq 0}\;{\cal C}_{0}^{\ell} \left[\;\frac{1}{\ell\Omega}\;{\cal C}_{\ell}^{0}[ \langle F\rangle] \;\right] \;+\; {\cal O}\left(\epsilon 
\epsilon_{\nu},\; \epsilon_{\nu}^{2}\right),
\label{eq:rcollop_new}
\end{equation}
where
\begin{equation}
{\cal C}_{0}^{0}[\langle F\rangle] \;\equiv\; \left\langle {\cal C}_{\epsilon}\left[\langle F\rangle \right] \right\rangle \;=\; \left\langle{\sf T}_{\epsilon}^{-1} {\cal C}[{\sf T}_{\epsilon}\langle F\rangle] \right\rangle
\label{eq:rcollop_00}
\end{equation}
is the lowest-order term in the closed transformed collision operator. 

The reduced collisional kinetic equation (\ref{eq:rcke_closed}), including the reduced collision operator 
(\ref{eq:rcollop_new}), is obtained from the collisional kinetic equation (\ref{eq:rcke}) as a result of two asymptotic expansions in powers of $\epsilon$ and $\epsilon_{\nu}$. In most practical applications, however, the zeroth-order expression (\ref{eq:rcollop_00}) is sufficient and, thus, we, henceforth, focus on applications of this lowest-order term in the development of reduced collisional kinetic theories. We note that the two small parameters $\epsilon_{B}$ and $\epsilon_{\nu}$ introduced above serve very different purposes: $\epsilon_{B} = \rho/L_{B}$ is involved in the elimination of fast gyro-motion time scale from the Vlasov operator and $\epsilon_{\nu} = \nu/\Omega = \rho/
\lambda_{\nu}$ is involved in the elimination of fast gyro-motion time scale from the collision operator. The ratio of these two dimensionless parameters introduces a new parameter $\Delta \equiv \epsilon_{B}/\epsilon_{\nu} = 
\lambda_{\nu}/L_{B}$ in the collisional kinetic theory of irreversible transport processes in nonuniform magnetized plasmas \cite{HH}. According to neoclassical collisional kinetic theory \cite{HH}, for example, the {\it classical} collisional regime is identified by the condition $\Delta \ll 1$, i.e., the collisional mean-free-path is much shorter than the magnetic nonuniformity length scale, so that magnetic nonuniformity can be ignored at the lowest order in 
$\epsilon_{B}$. In the {\it neoclassical} regime, on the other hand, we have $\Delta \geq 1$ (i.e., $\lambda_{\nu} \geq 
L_{B}$), while in the {\it collisionless} (long-mean-free-path) regime, we have $\Delta \gg 1$ (i.e., $\lambda_{\nu} \gg 
L_{B}$). In the latter two cases, the motion of a guiding-center particle explores a significant portion of the magnetic field nonuniformity before it encounters a collision.

\vspace*{0.2in}

\no
{\sf III. TRANSFORMED FOKKER-PLANCK COLLISION OPERATOR}

\vspace*{0.2in}

The collisional operator ${\cal C}$ has so far been treated as an arbitrary bilinear collision operator. We now focus our attention on an important 
class of bilinear collision operators used in plasma physics: the Fokker--Planck (FP) collision operators (see, for example, 
Refs.~\cite{Hinton,Rosenbluth,Rostoker}). The general form of the FP collision operator is expressed as
\begin{equation}
{\cal C}_{{\rm FP}}[f]({\bf x},{\bf p}) \;\equiv\; -\; \pd{}{p^{i}}\; \left[\; K^{i}({\bf x},{\bf p})\; f({\bf x},{\bf p}) \;-\; D^{ij}({\bf x},{\bf p})\; 
\pd{f}{p^{j}}({\bf x},{\bf p}) \;\right],
\label{eq:FP_op}
\end{equation}
where ${\bf z} \equiv ({\bf x},{\bf p})$ are phase-space coordinates for a test-particle of mass $m$ and charge $e$ (here, ${\bf x}$ denotes the particle position and ${\bf p} = m{\bf v}$ its kinetic momentum); the time dependence is suppressed for simplicity of notation and the momentum representation is used to facilitate the analysis that follows. The collision operator (\ref{eq:FP_op}) describes momentum-space scattering events in both unmagnetized plasmas \cite{Rosenbluth} and magnetized plasmas \cite{Rostoker,mfpc}: the three-component vector ${\bf K} = \sum^{\prime}\,{\bf K}[f']$ describes collisional momentum drag while the three-by-three symmetric matrix ${\sf D} = \sum^{\prime}\,{\sf D}[f']$ describes collisional momentum diffusion; the coefficients ${\bf K}[f']$ and ${\sf D}[f']$ are functionals of the {\it field-particle} distribution function $f'$. 

For small-angle-deflection Coulomb collisions in an unmagnetized plasma involving test-particle species $a$ and field-particle species $b$ \cite{HH}, for example, the coefficients ${\bf K}_{ab}[f_{b}]$ and ${\sf D}_{ab}[f_{b}]$ are expressed in terms of Rosenbluth potentials $(H_{ab},G_{ab})$ as
\begin{eqnarray}
{\bf K}_{ab}({\bf z}) & = & \pd{}{{\bf p}} \left( \Gamma_{ab}\;\frac{m_{a}}{m_{b}}\;\int d^{6}z^{\prime}\; f_{b}({\bf z}^{\prime})\;
\frac{\delta^{3}({\bf x}^{\prime} - {\bf x})}{|{\bf v} - {\bf v}^{\prime}|} \right) \;\equiv\; \pd{H_{ab}({\bf z})}{{\bf p}}, \nonumber \\ 
 &   & \\
{\sf D}_{ab}({\bf z}) & = & \frac{1}{2}\,\frac{\partial^{2}}{\partial{\bf p}\,\partial{\bf p}} \left( m_{a}^{2}\,\Gamma_{ab}\;
\int d^{6}z^{\prime}\; f_{b}({\bf z}^{\prime})\;\delta^{3}({\bf x}^{\prime} - {\bf x})\;|{\bf v} - {\bf v}^{\prime}| \right) \;\equiv\; 
\frac{1}{2}\; \frac{\partial^{2}G_{ab}({\bf z})}{\partial{\bf p}\,\partial{\bf p}}, \nonumber
\label{eq:Coulomb_op}
\end{eqnarray}
where $\Gamma_{ab} = 4\pi\,e_{a}^{2}e_{b}^{2}\;\ln\Lambda$ and $\partial_{{\bf p}}\bdot{\sf D}_{ab} = (m_{b}/m_{a})\,{\bf K}_{ab}$. Note that 
{\it magnetized} FP collision operators can also be expressed in terms of Rosenbluth potentials \cite{Rostoker,mfpc}. However, the unmagnetized FP collision operator (\ref{eq:FP_op}) can still be used to study collisional transport processes in magnetized plasmas if the characteristic gyroradius 
is larger than the Debye length for each plasma particle species (i.e., the Alfven speed is sub-luminal for each plasma particle species), which is well satisfied for electrons and ions in high-temperature tokamak plasmas. 

When the field-particle distribution is isotropic in momentum space \cite{Hinton}, such that the Rosenbluth potentials 
$H_{ab}({\bf z}) = H_{ab}({\bf x},\,p)$ and $G_{ab}({\bf z}) = G_{ab}({\bf x},\, p)$ depend only on the magnitude of the particle momentum, the FP coefficients become
\begin{equation} 
{\bf K}_{ab} \;=\; -\,\nu_{ab}\;{\bf p} \;\;\;{\rm and}\;\;\; {\sf D}_{ab} \;=\; D_{\|ab}\;
\frac{{\bf p}{\bf p}}{p^{2}} \;+\; D_{\bot ab} \left( {\bf I} \;-\; \frac{{\bf p}{\bf p}}{p^{2}} \right),
\label{eq:isotropic}
\end{equation}
where the collisional momentum-drag frequency is $\nu_{ab} = -\;(1/p)\,H_{ab}^{\prime}(p)$, the diffusion coefficients parallel and perpendicular to 
${\bf p}$, respectively, are $D_{\|ab} = (1/2)\,G_{ab}^{\prime\prime}(p)$ and $D_{\bot ab} = (1/2p)\,G_{ab}^{\prime}(p)$. The isotropic field-particle model will become useful later on when we investigate the effects of magnetic-field nonuniformity on guiding-center collisions. 

\vspace*{0.2in}

\no
{\bf A. Standard Transformation Procedure}

\vspace*{0.2in}

The standard transformation procedure for the FP collision operator (\ref{eq:FP_op}) is expressed in terms of the operator transformation ${\cal C}_{FP} \rightarrow {\cal C}_{TFP}$:
\begin{equation}
{\cal C}_{TFP}[F] \;\equiv\; -\;\frac{1}{{\cal J}}\;
\pd{}{Z^{\alpha}} \left[\; {\cal J} \left( {\cal K}^{\alpha} F \;-\; {\cal D}^{\alpha\beta}\; \pd{F}{Z^{\beta}}
\right) \;\right],
\label{eq:FP_standard}
\end{equation}
where ${\cal J}$ denotes the Jacobian of the general transformation ${\bf z} = ({\bf x},{\bf p}) \rightarrow {\bf Z} = {\cal T}\,{\bf z}$, the distribution function $F \equiv {\sf T}^{-1}f$ denotes the push-forward of the particle distribution function $f$ generated by the general 
transformation ${\cal T}$, and the new FP coefficients in Eq.~(\ref{eq:FP_standard}) are defined as
\begin{eqnarray} 
{\cal K}^{\alpha} & = & {\sf T}^{-1}\left( \pd{Z^{\alpha}}{p^{i}}\;K^{i}\right) \;=\; {\sf T}^{-1}\left( \pd{Z^{\alpha}}{p^{i}}\right)\;\cdot\;
{\sf T}^{-1}K^{i} \label{eq:K_standard} \\ 
{\cal D}^{\alpha\beta} & = & {\sf T}^{-1}\left( \pd{Z^{\alpha}}{p^{i}}\,D^{ij}\,\pd{Z^{\beta}}{p^{j}}\right) \;=\;
{\sf T}^{-1}\left( \pd{Z^{\alpha}}{p^{i}}\right)\;\cdot\;\left({\sf T}^{-1}D^{ij}\right)\;\cdot\; {\sf T}^{-1}\left(
\pd{Z^{\beta}}{p^{j}}\right), \label{eq:D_standard}
\end{eqnarray}
where we have used standard properties of the push-forward operator ${\sf T}^{-1}$: ${\sf T}^{-1}(f\;g) = ({\sf T}^{-1}f)\,({\sf T}^{-1}g)$. Note that, because the transformed FP collision operator (\ref{eq:FP_standard}) is written in divergence form, it is guaranteed to satisfy the particle conservation property.

Although the standard transformation procedure (\ref{eq:FP_standard}) is straightforward to apply for finite transformations \cite{Bernstein_Molvig} (e.g., transformation from Cartesian to cylindrical momentum coordinates), the explicit calculation of the transformation-matrix components 
${\sf T}_{\epsilon}^{-1}(\partial Z^{\alpha}_{\epsilon}/\partial p^{i})$ can be difficult to implement in practice for near-identity transformations on six-dimensional phase space, such as those contemplated in the present work. The difficulty resides in the fact that theterms ${\sf T}^{-1}_{\epsilon}(\partial Z_{\epsilon}^{\alpha}/\partial p^{i})$ in Eqs.~(\ref{eq:K_standard})-(\ref{eq:D_standard}) involve a double asymptotic expansion in powers of 
$\epsilon$ \cite{Balescu,CT,Auerbach}. To bypass this difficulty, we introduce a new formulation of the FP collision operator (\ref{eq:FP_op}) based on the use of the noncanonical Poisson bracket $\{\;,\;\}$ used in describing single-particle Hamiltonian dynamics.

\vspace*{0.2in}

\no
{\bf B. Poisson-Bracket Formulation}

\vspace*{0.2in}

First, we point out that the general FP collision operator (\ref{eq:FP_op}) can be written as
\begin{equation}
{\cal C}_{{\rm FP}}[f]({\bf z}) \;\equiv\; -\; \left\{ x^{i},\; \left[\;K^{i}({\bf z}) \;f({\bf z}) \;-\; D^{ij}({\bf z})\; \left\{ x^{j},\; f({\bf z}) \right\} 
\;\right] \;\right\},
\label{eq:FPop_PB}
\end{equation}
in terms of the {\it noncanonical} Poisson bracket \cite{Littlejohn_79}:
\begin{equation}
\{ f,\; g\}({\bf z}) \;=\; \pd{f({\bf z})}{{\bf x}}\;\vb{\cdot}\;\pd{g({\bf z})}{{\bf p}} \;-\; \pd{f({\bf z})}{{\bf p}}
\;\vb{\cdot}\; \pd{g({\bf z})}{{\bf x}} \;+\; \frac{e{\bf B}({\bf x})}{c}\;\vb{\cdot}\;\pd{f({\bf z})}{{\bf p}}\;\vb{\times}\;\pd{g({\bf z})}{{\bf p}},
\label{eq:PB}
\end{equation}
where $f$ and $g$ are two arbitrary phase-space functions and the magnetic field ${\bf B}$ appears in Eq.~(\ref{eq:PB}) because ${\bf p} = m{\bf v}$ is not the canonical momentum. 

Next, using the transformed Poisson bracket (\ref{eq:r_PB}), the transformed FP collision operator is obtained from Eq.~(\ref{eq:rcoll_op}) as
\begin{equation}
{\cal C}_{\epsilon{\rm FP}}[F]({\bf Z}) \;=\; -\; \left\{ X_{\epsilon}^{i}({\bf Z}),\; \left[\; K_{\epsilon}^{i}({\bf Z})\; F({\bf Z}) \;-\; D_{\epsilon}^{ij}({\bf Z})\; \left\{ X_{\epsilon}^{j}({\bf Z}),\; F({\bf Z}) \right\}_{\epsilon} \,\right] \;\right\}_{\epsilon},
\label{eq:epsilon_linFP}
\end{equation}
where the transformed FP coefficients are $K_{\epsilon}^{i}({\bf Z}) = {\sf T}_{\epsilon}^{-1}K^{i}({\bf Z})$ and 
$D_{\epsilon}^{ij}({\bf Z}) = {\sf T}_{\epsilon}^{-1}D^{ij}({\bf Z})$, and $X_{\epsilon}^{i}({\bf Z}) = 
{\sf T}_{\epsilon}^{-1}x^{i}({\bf Z}) \equiv x^{i}$ represents the particle position ${\bf x}$ expressed as a function of the transformed phase-space coordinates ${\bf x} = {\bf X}_{\epsilon}({\bf Z})$, and ${\bf X}$ represents the transformed spatial coordinate. 

We now point out that a general Poisson bracket in six-dimensional phase space (with coordinates $Z^{\alpha}$) can be expressed in terms of a six-by-six antisymmetric Poisson matrix $J_{\epsilon}^{\alpha\beta} = \{ Z^{\alpha},\; Z^{\beta}\}_{\epsilon}$ and the Jacobian ${\cal J}$ as a phase-space divergence
\[ \{ F,\; G \}_{\epsilon} \;=\; \pd{F}{Z^{\alpha}}\;J_{\epsilon}^{\alpha\beta}\;\pd{G}{Z^{\beta}} \;\equiv\; -\;
\frac{1}{{\cal J}}\;\pd{}{Z^{\alpha}} \left( {\cal J}\;\pd{F}{Z^{\beta}}\;J_{\epsilon}^{\alpha\beta}\;G \right), \]
where the functions $F$ and $G$ are arbitrary functions on the transformed phase space, while the Jacobian ${\cal J}$ and the anti-symmetric Poisson matrix $J_{\epsilon}^{\alpha\beta}$ satisfy the Liouville identities $\partial_{\alpha}({\cal J}\,J_{\epsilon}^{\alpha\beta}) = 0$ (for $\beta = 1,
...6$). With this Poisson-bracket formulation, we may rewrite the Fokker-Planck collision operator (\ref{eq:epsilon_linFP}) in the same phase-space divergence form as Eq.~(\ref{eq:FP_standard}):
\begin{equation}
{\cal C}_{\epsilon{\rm FP}}[F]({\bf Z}) \;\equiv\; -\;\frac{1}{{\cal J}} \pd{}{Z^{\alpha}} \left[\; {\cal J} \left( {\cal K}_{\epsilon}^{\alpha}\;F \;-\; {\cal D}_{\epsilon}^{\alpha\beta}\; \pd{F}{Z^{\beta}} \right) \;\right],
\label{eq:epsilon_divFP}
\end{equation}
where the phase-space collisional drag vector is defined as
\begin{equation}
{\cal K}_{\epsilon}^{\alpha} \;=\; {\bf K}_{\epsilon}\bdot\partial_{\sigma}{\bf X}_{\epsilon}\;J_{\epsilon}^{\sigma\alpha} \;\equiv\; {\bf K}_{\epsilon}\bdot\vb{\Delta}_{\epsilon}^{\alpha},
\label{eq:epsilon_drift}
\end{equation}
while the phase-space collisional diffusion tensor is defined as
\begin{equation}
{\cal D}_{\epsilon}^{\alpha\beta} \;=\; -\;J_{\epsilon}^{\alpha\mu}\,\partial_{\mu}{\bf X}_{\epsilon}\bdot
{\sf D}_{\epsilon}\bdot\partial_{\nu}{\bf X}_{\epsilon}\;J_{\epsilon}^{\nu\beta} \;\equiv\; 
\vb{\Delta}_{\epsilon}^{\alpha}\bdot{\sf D}_{\epsilon}\bdot\vb{\Delta}_{\epsilon}^{\beta},
\label{eq:epsilon_diffusion}
\end{equation}
which is also symmetric since $D^{ij}_{\epsilon}$ is symmetric. In Eqs.~(\ref{eq:epsilon_drift}) and (\ref{eq:epsilon_diffusion}), we have introduced the vector-valued components $\vb{\Delta}_{\epsilon}^{\alpha}$ defined as
\begin{equation}
\vb{\Delta}_{\epsilon}^{\alpha} \;=\; \left\{ {\bf X}_{\epsilon},\; Z^{\alpha} \right\}_{\epsilon} \;=\; -\;
J^{\alpha\beta}\;\pd{{\bf X}_{\epsilon}}{Z^{\beta}},
\label{eq:Delta_alpha}
\end{equation}
which offer computational advantages over the standard transformation approach (described above) since asymptotic expressions for the antisymmetric Poisson matrix $J_{\epsilon}^{\alpha\beta}$ and ${\bf X}_{\epsilon}$ are readily available for the guiding-center \cite{gc,Brizard_95} and bounce-center \cite{bc} near-identity phase-space transformations. 

In the next Section, we derive expressions for the guiding-center Fokker-Planck collision operator based on the general phase-space divergence form (\ref{eq:epsilon_divFP}), with coefficients given by Eqs.~(\ref{eq:epsilon_drift}) and
(\ref{eq:epsilon_diffusion}).

\vspace*{0.2in}

\no
{\sf IV. REDUCED GUIDING-CENTER FOKKER-PLANCK OPERATOR}

\vspace*{0.2in}

As an application of the Poisson-bracket formalism presented in Sec.~III.B, we consider the transformation of a general FP collision operator under the guiding-center transformation \cite{gc}: ${\bf z} = ({\bf x},{\bf p}) \rightarrow {\bf Z} = ({\bf X},{\cal E};\mu,\theta)$, where ${\bf X}$ denotes the guiding-center position, ${\cal E}$ denotes the guiding-center kinetic energy (${\cal E} = p_{\|}^{2}/2m + \mu B$), $\mu$ denotes the guiding-center magnetic moment, and $\theta$ denotes the guiding-center gyroangle. The choice of the actual velocity-space guiding-center coordinates used in the guiding-center FP collision operator can vary depending on its applications. Here, we use the $({\cal E},\mu)$ coordinates because of the simplicity of the derived guiding-center FP collision operator. Future work will consider replacing the magnetic moment $\mu$ with the guiding-center's pitch angle 
$\xi = \arccos(p_{\|}/p)$.

\vspace*{0.2in}

\no
{\bf A. Nonuniform Magnetic Field}

\vspace*{0.2in}

In the collisionless long-mean-free-path regime $\Delta = \epsilon_{B}/\epsilon_{\nu} = \lambda_{\nu}/L_{B} \gg 1$ (where the collisional mean-free-path $\lambda_{\nu}$ is much longer than the magnetic nonuniformity length scale $L_{B}$), the guiding-center motion of a charged particle samples a significant portion of the magnetic-field nonuniformity before it encounters a collision and, thus, magnetic field nonuniformity becomes important when describing small-angle collisions.

According to the guiding-center transformation \cite{gc,Brizard_95}, the particle position 
\begin{equation}
{\bf x} \;=\; {\bf X} \;+\; \vb{\rho}_{\epsilon} \;\equiv\; {\bf X}_{\epsilon}
\label{eq:x_X_gyro}
\end{equation}
is expressed in terms of the guiding-center position ${\bf X}$ and the gyroradius vector $\vb{\rho}_{\epsilon}$ (whose expression includes corrections due to magnetic-field nonuniformity); the dimensionless parameter $\epsilon = \epsilon_{B}$ represents the magnetic field nonuniformity. The expression for the gyroradius vector (up to order 
$\epsilon$)
\begin{equation}
\vb{\rho}_{\epsilon}({\bf Z}) \;=\; \vb{\rho}_{0} \;-\; \epsilon \left( G_{2}^{{\bf X}} \;+\; \frac{1}{2}\;G_{1}\cdot d\,\vb{\rho}_{0} \right) \;+\; 
\cdots
\label{eq:rho_epsilon}
\end{equation}
is given in terms of the components of the first and second order generating vector fields $(G_{1}^{\alpha}, G_{2}^{\alpha})$ for the guiding-center transformation \cite{gc,Brizard_95} and contains the lowest-order gyroradius vector $\vb{\rho}_{0} = -\,G_{1}^{{\bf X}}$ as well as the first-order correction $\vb{\rho}_{1} = -\,G_{2}^{{\bf X}} - \frac{1}{2}\;G_{1}\cdot d\,\vb{\rho}_{0}$; an explicit expression for the gyroradius first-order correction $\vb{\rho}_{1}$ is not needed in this Section and is given in Appendix A [see Eq.~(\ref{eq:rho_1})].   

The Jacobian ${\cal J} = m\,B_{\|}^{*}/|v_{\|}|$ for the guiding-center transformation \cite{gc} is defined in terms of the guiding-center phase-space function $B_{\|}^{*}({\bf X},{\cal E},\mu) = \bhat\bdot{\bf B}^{*}$ derived from the generalized magnetic field ${\bf B}^{*} = {\bf B} + \epsilon\,(cp_{\|}/e)\;\nabla\btimes\wh{{\sf b}}$, where $|v_{\|}| = |p_{\|}|/m = \sqrt{(2/m)\,({\cal E} - \mu\,B)}$. Next, the guiding-center Poisson bracket of two arbitrary functions $F$ and $G$ of $({\bf X},{\cal E},\mu,\theta)$ is
\begin{eqnarray} 
\{ F,\; G \}_{{\rm gc}} & = & \frac{\Omega}{B}\; \left[\; \pd{F}{\theta} \left( \pd{G}{\mu} + B\,\pd{G}{{\cal E}}\right) 
\;-\; \left( \pd{F}{\mu} + B\,\pd{F}{{\cal E}} \right) \pd{G}{\theta} \;\right] \nonumber \\ 
 &   &\mbox{}+\; {\bf v}_{{\rm gc}}\bdot\left( \nabla^{*} F\,\pd{G}{{\cal E}} \;-\; \pd{F}{{\cal E}}\,\nabla^{*} G 
\right) \;-\; \frac{c\wh{{\bf b}}}{eB_{\|}^{*}}\bdot\nabla^{*} F\btimes\nabla^{*} G, 
\label{eq:gc_PB}
\end{eqnarray}
where $\nabla^{*}$ is a gradient operator defined in Appendix A [see Eq.~(\ref{eq:grad_rho})] and the gyro-averaged guiding-center velocity is
\begin{equation}
{\bf v}_{{\rm gc}} \;=\; v_{\|}\;\bhat \;+\; \left. \left. \frac{\bhat}{m\Omega_{\|}^{*}}\btimes\right( \mu\;\nabla B 
\;+\; mv_{\|}^{2}\;\bhat\bdot\nabla\bhat \right),
\label{eq:gc_velocity}
\end{equation}
with $\Omega_{\|}^{*} = eB_{\|}^{*}/(mc) = \Omega + v_{\|}\;\bhat\bdot\nabla\btimes\bhat$. Lastly, the particle momentum ${\bf p}_{\epsilon}({\bf X},{\cal E},\mu,\theta) = {\sf T}_{\epsilon}^{-1}{\bf p}$ is written in guiding-center coordinates as
\begin{equation}
{\bf p}_{\epsilon} \;=\; m\;\frac{d_{\epsilon}{\bf X}_{\epsilon}}{dt} \;=\; m \left[\; {\bf v}_{{\rm gc}} \;+\; \left( \Omega\;\pd{}{\theta} \;+\; 
{\bf v}_{{\rm gc}}\bdot\nabla^{*}\right)\vb{\rho}_{\epsilon} \;\right].
\label{eq:p_epsilon}
\end{equation}
An explicit expression for ${\bf p}_{\epsilon}$ is given in Appendix A [see Eq.~(\ref{eq:Peps_full})]; here, we simply note that ${\bf p}_{\epsilon}$ satisfies the following properties
\begin{equation} 
|{\bf p}_{\epsilon}|^{2} \;=\; |{\bf p}|^{2} \;=\; 2m\, {\cal E} \;\;\;{\rm and}\;\;\; \langle {\bf p}_{\epsilon}\rangle \;=\; m\,{\bf v}_{{\rm gc}},
\label{eq:Peps_prop}
\end{equation}
up to first order in $\epsilon$.

\vspace*{0.2in}

\no
{\bf B. Guiding-center Fokker-Planck Collision Operator}

\vspace*{0.2in}

We now proceed with obtaining general expressions for the phase-space collisional drag vector (\ref{eq:epsilon_drift}) and the phase-space collisional diffusion tensor (\ref{eq:epsilon_diffusion}) associated with the guiding-center transformation $({\bf x},{\bf p}) \rightarrow ({\bf X},{\cal E},\mu,\theta)$. First, using the definition (\ref{eq:Delta_alpha}) for $\vb{\Delta}_{{\rm gc}}^{\alpha}$ and the guiding-center Poisson bracket 
(\ref{eq:gc_PB}), the guiding-center vector-valued components of $\vb{\Delta}_{{\rm gc}}^{\alpha} = \{ {\bf X}_{\epsilon},\; Z^{\alpha}\}_{{\rm gc}}$ are
\begin{eqnarray}
\vb{\Delta}_{{\rm gc}}^{i} & = & -\;\frac{c\,\epsilon^{ijk}}{eB_{\|}^{*}}\;b_{j}\,\partial_{k}^{*}{\bf X}_{\epsilon} 
\;-\; v_{{\rm gc}}^{i}\;\pd{\vb{\rho}_{\epsilon}}{{\cal E}}, \label{eq:Delta_i} \\
\vb{\Delta}_{{\rm gc}}^{{\cal E}} & = & {\bf v}_{{\rm gc}} \;+\; \left\{ \vb{\rho}_{\epsilon},\; 
{\cal E} \right\}_{{\rm gc}} \;=\; \frac{{\bf p}_{\epsilon}}{m}, 
\label{eq:Delta_E} \\
\vb{\Delta}_{{\rm gc}}^{\mu} & = & \frac{\Omega}{B}\;\pd{\vb{\rho}_{\epsilon}}{\theta}, 
\label{eq:Delta_mu}
\end{eqnarray}
and we have ignored the component-vector $\vb{\Delta}_{{\rm gc}}^{\theta}$ since it is not needed in evaluating the lowest-order guiding-center Fokker-Planck collision operator (\ref{eq:rcollop_00}). 

With these component-vectors (a greek-letter index, henceforth, excludes the gyroangle component), the guiding-center phase-space collisional drag vector and the guiding-center phase-space collisional diffusion tensor have the following guiding-center phase-space components
\begin{equation}
{\cal K}_{{\rm gc}}^{\alpha} \;=\; {\bf K}_{\epsilon}\bdot\left\{ {\bf X}_{\epsilon},\; Z^{\alpha} \right\}_{{\rm gc}} \;\;\;{\rm and}\;\;\; 
{\cal D}_{{\rm gc}}^{\alpha\beta} \;=\; -\; \left\{ Z^{\alpha},\; {\bf X}_{\epsilon}\right\}_{{\rm gc}}\bdot{\sf D}_{\epsilon}\bdot\left\{ {\bf X}_{\epsilon},\; Z^{\beta} \right\}_{{\rm gc}}. 
\label{eq:KD_gc}
\end{equation}
The lowest-order guiding-center Fokker-Planck collision operator is, thus, expressed in terms of the gyroangle-averaged guiding-center distribution 
$\langle F\rangle$ as
\begin{equation}
{\cal C}_{{\rm gcFP}}[\langle F\rangle] \;=\; -\;\frac{1}{{\cal J}}\;\pd{}{Z^{\alpha}} \left[\; {\cal J} \left( \langle 
K_{{\rm gc}}^{\alpha}\rangle\;\langle F\rangle \;-\; \langle D_{{\rm gc}}^{\alpha\beta}\rangle\;
\pd{\langle F\rangle}{Z^{\beta}} \right) \;\right],
\label{eq:gcFP_general}
\end{equation}
where expressions for the guiding-center FP coefficients $\langle K_{{\rm gc}}^{\alpha}\rangle$ and $\langle 
D_{{\rm gc}}^{\alpha\beta}\rangle$ are given with terms up to order $\epsilon_{B}$. Note that the guiding-center Fokker-Planck collision operator now contains spatial derivatives and, therefore, Eq.~(\ref{eq:gcFP_general}) describes spatial collisional drag and spatial collisional diffusion.

\vspace*{0.1in}

\no
{\it 1. Guiding-center collisional drag vector}

\vspace*{0.1in}

The gyroangle-averaged spatial components of the guiding-center collisional drag vector needed for the lowest-order guiding-center FP collision operator 
(\ref{eq:rcollop_00}) are
\begin{equation}
\left\langle K_{{\rm gc}}^{{\bf X}}\right\rangle \;=\; \langle{\bf K}_{\epsilon}\rangle\btimes
\frac{\bhat}{m\Omega_{\|}^{*}} \;+\; \cdots,
\label{eq:gcK_spatial}
\end{equation}
where the omitted terms are of second order in $\epsilon_{B}$. The gyroangle-averaged energy and magnetic-moment components, on the other hand, are
\begin{eqnarray}
\left\langle K_{{\rm gc}}^{{\cal E}}\right\rangle & = & \left\langle {\bf K}_{\epsilon}\bdot\frac{{\bf p}_{\epsilon}}{m} \right\rangle, \label{eq:Kgc_E} \\
 &  & \nonumber \\
\left\langle K_{{\rm gc}}^{\mu}\right\rangle & = & \left\langle {\bf K}_{\epsilon}\bdot \frac{\Omega}{B}\;\pd{\vb{\rho}_{\epsilon}}{\theta} \right\rangle.
\label{eq:Kgc_mu}
\end{eqnarray}
We note that the components (\ref{eq:Kgc_E})-(\ref{eq:Kgc_mu}) of the guiding-center phase-space collisional drag vector are non-vanishing in the 
absence of magnetic nonuniformity and, thus, magnetic nonuniformity adds only small first-order corrections to these components. We expect the spatial components (\ref{eq:gcK_spatial}) of the guiding-center phase-space collisional drag vector, on the other hand, to vanish in the absence of magnetic nonuniformity as will be demonstrated when we apply the isotropic-field-particle model (\ref{eq:isotropic}) to Eq.~(\ref{eq:gcK_spatial}).

\vspace*{0.1in}

\no
{\it 2. Guiding-center collisional diffusion tensor}

\vspace*{0.1in}

The gyroangle-averaged spatial components of the guiding-center diffusion tensor are
\begin{equation}
\left\langle D_{{\rm gc}}^{{\bf X}{\bf X}}\right\rangle \;=\; -\;\frac{\bhat}{m\Omega_{\|}^{*}}\btimes\langle 
{\sf D}_{\epsilon}\rangle\btimes\frac{\bhat}{m\Omega_{\|}^{*}} \;+\; \cdots,
\label{eq:Dgc_ij}
\end{equation}
where higher-order terms are omitted since spatial diffusion involves second-order spatial gradients already. The gyroangle-averaged energy and magnetic-moment diffusion coefficients are
\begin{eqnarray}
\left\langle D_{{\rm gc}}^{{\cal E}{\cal E}}\right\rangle & = & \left\langle \frac{{\bf p}_{\epsilon}}{m}\bdot
{\sf D}_{\epsilon}\bdot\frac{{\bf p}_{\epsilon}}{m} \right\rangle, \label{eq:Dgc_EE} \\
\left\langle D_{{\rm gc}}^{\mu{\cal E}} \right\rangle & = & \left\langle \frac{\Omega}{B}\;\pd{\vb{\rho}_{\epsilon}}{\theta}\bdot{\sf D}_{\epsilon}\bdot\frac{{\bf p}_{\epsilon}}{m} \right\rangle, \label{eq:Dgc_Emu} \\
\left\langle D_{{\rm gc}}^{\mu\mu}\right\rangle & = & \left\langle \frac{\Omega}{B}\;\pd{\vb{\rho}_{\epsilon}}{\theta}\bdot{\sf D}_{\epsilon}\bdot
\frac{\Omega}{B}\;\pd{\vb{\rho}_{\epsilon}}{\theta}\right\rangle, \label{eq:Dgc_mumu}
\end{eqnarray}
and the gyroangle-averaged mixed-component {\it spatial} diffusion coefficients are
\begin{eqnarray}
\left\langle D_{{\rm gc}}^{{\bf X}{\cal E}} \right\rangle & = & \frac{c\bhat}{eB_{\|}^{*}}\btimes\left\langle 
{\sf D}_{\epsilon}\bdot \frac{{\bf p}_{\epsilon}}{m} \right\rangle, \label{eq:Dgc_iE} \\
\left\langle D_{{\rm gc}}^{{\bf X}\mu} \right\rangle & = & \frac{c\bhat}{eB_{\|}^{*}}\btimes\left\langle 
{\sf D}_{\epsilon}\bdot \frac{\Omega}{B}\;\pd{\vb{\rho}_{\epsilon}}{\theta} \right\rangle.
\label{eq:Dgc_imu} 
\end{eqnarray}
Once again we note that the components (\ref{eq:Dgc_ij})-(\ref{eq:Dgc_mumu}) of the guiding-center phase-space collisional diffusion tensor are non-vanishing in the absence of magnetic nonuniformity and, thus, magnetic nonuniformity only adds small corrections to these components. We expect the spatial components (\ref{eq:Dgc_iE})-(\ref{eq:Dgc_imu}) of the guiding-center phase-space collisional diffusion tensor, on the other hand, to vanish in the absence of magnetic nonuniformity.

\vspace*{0.2in}

\no
{\bf C. Isotropic Field-Particle Model}

\vspace*{0.2in}

To obtain a specific expression for a guiding-center FP collision operator (\ref{eq:gcFP_general}) and investigate how a nonuniform magnetic field affects the form of the FP collision operator, we now consider the case of an isotropic field-particle (IFP) distribution [see Eq.~(\ref{eq:isotropic})], for which the transformed collisional drag vector and the collisional diffusion tensor are
\begin{eqnarray}
{\bf K}_{\epsilon} & = & \left. \left. {\sf T}_{\epsilon}^{-1}\right(-\nu\,{\bf p}\right) \;=\; -\;\nu_{\epsilon}\;
{\bf p}_{\epsilon}, \label{eq:iso_K} \\
{\sf D}_{\epsilon} & = & \left. \left. {\sf T}_{\epsilon}^{-1}\right[\; D_{\bot}\;{\bf I} \;+\; \left( D_{\|} - D_{\bot}\right)\; \frac{{\bf p}{\bf p}}{p^{2}} \;\right] \;=\; D_{\bot\epsilon}\;{\bf I} \;+\; \left( D_{\|\epsilon} \;-\; 
D_{\bot\epsilon} \right) \frac{{\bf p}_{\epsilon}{\bf p}_{\epsilon}}{2m\,{\cal E}},
\label{eq:iso_D}
\end{eqnarray}
where $\nu_{\epsilon}$, $D_{\|\epsilon}$, and $D_{\bot\epsilon}$ are calculated from Rosenbluth potentials and are assumed to depend only on the guiding-center energy ${\cal E}$ (to lowest order in spatial nonuniformity) for simplicity of presentation. From Eqs.~(\ref{eq:iso_K})-(\ref{eq:iso_D}) and (\ref{eq:Peps_prop}), we obtain the IFP identities
\begin{equation}
{\bf K}_{\epsilon}\bdot{\bf p}_{\epsilon} \;=\; -\;\nu_{\epsilon}\;2m{\cal E} \;\;\; {\rm and}\;\;\;
{\bf p}_{\epsilon}\bdot{\sf D}_{\epsilon}\bdot{\bf p}_{\epsilon} \;=\; {\bf p}_{\epsilon}\bdot(D_{\|\epsilon}\;
{\bf p}_{\epsilon}) \;=\; D_{\|\epsilon}\;2m{\cal E},
\label{eq:IFP_defs}
\end{equation}
and the gyroangle-averaged expressions
\begin{eqnarray} 
\left\langle {\bf K}_{\epsilon}\right\rangle & = & -\,\nu_{\epsilon}\;m\,{\bf v}_{{\rm gc}}, \nonumber \\
\left\langle {\sf D}_{\epsilon}\bdot{\bf p}_{\epsilon}\right\rangle & = & {\cal D}_{{\cal E}}\;m\,{\bf v}_{{\rm gc}},
\label{eq:IFP_average} \\
\wh{{\sf b}}\btimes\langle {\sf D}_{\epsilon}\rangle\btimes\wh{{\sf b}} & = & -\;\left( D_{\bot\epsilon} \;+\; 
m\,{\cal D}_{\mu}\,B \right) {\bf I}_{\bot}, \nonumber
\end{eqnarray}
where ${\bf I}_{\bot} = {\bf I} - \bhat\,\bhat$ and we introduced the convenient notation
\begin{equation} 
{\cal D}_{{\cal E}} \;\equiv\; \frac{D_{\|\epsilon}}{m} \;\;\;{\rm and}\;\;\; {\cal D}_{\mu} \;=\; \mu\; \left( 
\frac{D_{\|\epsilon} - D_{\bot\epsilon}}{2m\;{\cal E}} \right).
\label{eq:IFP_notation}
\end{equation}
We now substitute Eqs.~(\ref{eq:iso_K})-(\ref{eq:iso_D}) and (\ref{eq:IFP_average})-(\ref{eq:IFP_notation}) to obtain explicit expressions for the guiding-center FP coefficients (\ref{eq:gcK_spatial})-(\ref{eq:Dgc_imu}). 

First, using the gyroangle-averaged expression for ${\sf D}_{\epsilon}$, the guiding-center FP spatial diffusion tensor (\ref{eq:Dgc_ij}) becomes (to lowest order in $\epsilon_{B}$)
\begin{equation} 
\langle D_{{\rm gc}}^{{\bf X}{\bf X}}\rangle \;=\; \frac{B}{m\Omega^{2}} \left( {\cal D}_{\mu} \;+\; 
\frac{D_{\bot\epsilon}}{mB} \right)\;{\bf I}_{\bot},
\label{eq:ifpD_space}
\end{equation}
which is identical to the guiding-center spatial diffusion coefficient derived by Xu and Rosenbluth \cite{XR}.
Using the identities (\ref{eq:IFP_defs}), we also easily find, from Eqs.~(\ref{eq:Kgc_E}) and (\ref{eq:Dgc_EE}), the following guiding-center FP coefficients
\begin{equation}
\left( \begin{array}{c}
\left\langle K_{{\rm gc}}^{{\cal E}}\right\rangle \\
\left\langle D_{{\rm gc}}^{{\cal E}{\cal E}}\right\rangle
\end{array} \right) \;=\; \left( \begin{array}{c}
-\;\nu_{\epsilon} \\
{\cal D}_{{\cal E}}
\end{array} \right) \;2\,{\cal E},
\label{eq:KD_E}
\end{equation}
which are unaffected by magnetic-field nonuniformity  as a result of the fact that the guiding-center and particle kinetic energies are equal to each other up to first order in $\epsilon_{B}$ [see Eq.~(\ref{eq:gc_E})]. 

Next, using the gyroangle-averaged expressions (\ref{eq:IFP_average}), the guiding-center FP spatial drag coefficient 
(\ref{eq:gcK_spatial}) and the guiding-center FP off-diagonal energy coefficient (\ref{eq:Dgc_iE}) become
\begin{equation}
\left( \begin{array}{c}
\left\langle K_{{\rm gc}}^{{\bf X}}\right\rangle \\
\left\langle D_{{\rm gc}}^{{\bf X}{\cal E}} \right\rangle
\end{array} \right) \;=\; \epsilon_{B} \left( \begin{array}{c}
\nu_{\epsilon} \\
{\cal D}_{{\cal E}}
\end{array} \right) \frac{\bhat}{\Omega_{\|}^{*}}\btimes{\bf v}_{{\rm gc}}.
\label{eq:IFP_KDspace}
\end{equation}
For the remaining guiding-center FP coefficients [(\ref{eq:Kgc_mu}), (\ref{eq:Dgc_Emu}), (\ref{eq:Dgc_mumu}), and
(\ref{eq:Dgc_imu})], we find
\begin{eqnarray}
\left( \begin{array}{c}
\left\langle K_{{\rm gc}}^{\mu}\right\rangle \\
\left\langle D_{{\rm gc}}^{\mu{\cal E}} \right\rangle
\end{array} \right) & = & (2 - \epsilon_{B}\,\lambda_{{\rm gc}})\;\left( \begin{array}{c}
-\;\nu_{\epsilon} \\
{\cal D}_{{\cal E}}
\end{array} \right)\,\mu, \label{eq:IFP_KDgc_mu} \\
 &  & \nonumber \\
\left\langle D_{{\rm gc}}^{\mu\mu}\right\rangle & = & (1 - \epsilon_{B}\,\lambda_{{\rm gc}})\; 2\mu\,\left( 
2{\cal D}_{\mu} \;+\; \frac{D_{\bot\epsilon}}{m\,B} \;\right), \label{eq:IFP_Dgc_mumu}
\end{eqnarray}
where $\lambda_{{\rm gc}} = (v_{\|}/\Omega)\,\bhat\bdot\nabla\btimes\bhat$ denotes the guiding-center {\it vorticity} parameter, and
\begin{equation}
\left\langle D_{{\rm gc}}^{{\bf X}\mu} \right\rangle \;=\; \epsilon_{B}\,{\cal D}_{\mu}\;\frac{\wh{{\sf b}}}{\Omega_{\|}^{*}}
\btimes{\bf v}_{{\rm gc}}, \label{eq:IFP_Dgc_imu} 
\end{equation}
where gyrorangle averages in Eqs.~(\ref{eq:IFP_KDspace})-(\ref{eq:IFP_Dgc_imu}) were computed using formulas found in Appendix A [see 
Eqs.~(\ref{muE_ifp})-(\ref{muspace_ifp})]. As was indicated above, the guiding-center FP spatial drag and diffusion coefficients (\ref{eq:IFP_KDspace}) and (\ref{eq:IFP_Dgc_imu}) vanish in the limit of a uniform magnetic field (i.e., $\epsilon_{B} = 0$), while the guiding-center FP velocity-space drag and diffusion coefficients (\ref{eq:IFP_KDgc_mu})-(\ref{eq:IFP_Dgc_mumu}) exhibit first-order corrections due to magnetic-field nonuniformity. 

\vspace*{0.2in}

\no
{\bf D. Guiding-center Ordering and Beyond}

\vspace*{0.2in}

\no
{\it 1. Guiding-center ordering}

\vspace*{0.1in}

In the standard guiding-center ordering, where $\epsilon_{B} \ll 1$, the first-order corrections in the IFP Fokker-Planck coefficients (\ref{eq:IFP_KDgc_mu})-(\ref{eq:IFP_Dgc_mumu}) are indeed quite small and can, thus, be neglected. The non-vanishing contributions (\ref{eq:IFP_KDspace}) and (\ref{eq:IFP_Dgc_imu}) associated with spatial drag and diffusion, on the other hand, yield terms of order comparable to the IFP spatial diffusion term (\ref{eq:ifpD_space}) if magnetic and plasma nonuniformity length scales are comparable (i.e., $\rho\,|\nabla_{\bot}\ln\langle F\rangle| \sim \epsilon_{B}$) as might be expected in compact high-temperature tokamak plasmas.

\vspace*{0.1in}

\no
{\it 2. Background electric field}

\vspace*{0.1in}

We now discuss how background electric fields might be included in the IFP model 
(\ref{eq:ifpD_space})-(\ref{eq:IFP_Dgc_imu}), with separate orderings $\epsilon_{E}$ and $\epsilon_{E}^{\prime}$  associated with the background $E \times B$ velocity and its derivatives \cite{Brizard_95,HC}, respectively. 

First, the term $\bhat\btimes{\bf v}_{{\rm gc}}$ in the IFP spatial components (\ref{eq:IFP_KDspace}) and 
(\ref{eq:IFP_Dgc_imu}) is replaced with $\bhat\btimes(\epsilon_{B}\,{\bf v}_{{\rm gc}} + \epsilon_{E}\,
{\bf v}_{{\rm E}})$, where $\epsilon_{E}$ denotes the ratio of the $E \times B$ velocity to the characteristic thermal speed (which is small in the core region of tokamak plasmas \cite{Brizard_95} but is larger in the edge region of tokamak plasmas with sheared flows \cite{HC}). Next, the term $\lambda_{{\rm gc}}$ in the velocity-space components 
(\ref{eq:IFP_KDgc_mu})-(\ref{eq:IFP_Dgc_mumu}) is replaced with $\epsilon_{B}\,\lambda_{{\rm gc}} + \epsilon_{E}^{\prime}
\,(\bhat/\Omega)\bdot\nabla\btimes{\bf v}_{{\rm E}}$, where $\epsilon_{E}^{\prime} = \rho/L_{E}$ denotes the ratio of the characteristic gyro-radius to the sheared-flow length scale. Here, the second term describes the orbit-squeezing effect of edge electric fields \cite{KCS}, which might lead to important modifications of collisional transport theory \cite{HC,KCS,SZ}. Future work will investigate the role played by background electric fields, which are relevant to the numerical gyrokinetic simulations of edge plasmas \cite{Xu_03}.

\vspace*{0.2in}

\no
{\sf V. SUMMARY}

\vspace*{0.2in}

In summary, we have derived a general reduced Fokker-Planck kinetic equation by Lie-transform methods. The collisional kinetic equation (\ref{eq:rcke_closed}) was obtained as a result of two independent asymptotic expansions: the first expansion (with small parameter $\epsilon = \omega/\Omega$) was associated with the elimination of the fast time scale from the Vlasov operator; while the second expansion (with small parameter $\epsilon_{\nu} = \nu/\Omega$) was associated with the elimination of the fast time scale from the transformed collision operator. 

As an application of the dynamical reduction formalism for bilinear collision operators, we derived a general guiding-center Fokker-Planck collision operator (\ref{eq:gcFP_general}), with coefficients given in compact form by 
Eqs.~(\ref{eq:gcK_spatial})-(\ref{eq:Dgc_imu}). In order to show explicitly how magnetic-field nonuniformity affects the form of the guiding-center FP collision operator, we also made use of the simple isotropic-field-particle model 
(\ref{eq:isotropic}) and obtained explicit expressions for the guiding-center FP coefficients (\ref{eq:ifpD_space})-(
\ref{eq:IFP_Dgc_imu}). In this simplified model, the effects of magnetic-field nonuniformity are represented by the gyroangle-averaged guiding-center velocity ${\bf v}_{{\rm gc}}$ and the guiding-center vorticity parameter 
$\lambda_{{\rm gc}}$.

Future work will consider applications of the formalism leading to the reduced guiding-center Fokker-Planck collision operator (\ref{eq:gcFP_general}) and the inclusion of effects associated with background sheared $E \times B$ velocities.

\vspace*{0.2in}

\no
{\sf ACKNOWLEDGMENTS}

\vspace*{0.2in}

This work was supported by the U.S.~Department of Energy under contract No. DE-AC03-76SF00098 and the National Science Foundation under grant No. DMS-0317339.

\vspace*{0.2in}

\setcounter{equation}{0}
\renewcommand{\theequation}{A.\arabic{equation}}

\no
{\sf Appendix A. First-Order Guiding-Center Corrections}

\vspace*{0.2in}

The guiding-center transformation $({\bf x},{\bf p}) \rightarrow ({\bf X},{\cal E},\mu,\theta)$ was originally derived by Littlejohn \cite{gc} by Lie-transform methods in the form of asymptotic expansions $Z^{\alpha}_{{\rm gc}} = 
Z_{0}^{\alpha} + \epsilon\,G_{1}^{\alpha} + \cdots$, where the components of the first-order generating vector field are
\begin{eqnarray}
G_{1}^{{\bf X}} & = & -\;\vb{\rho}_{0} \;=\; -\;\frac{mc}{e}\;\sqrt{\frac{2\,\mu}{m B(\epsilon{\bf X})}}\;
\wh{\rho}(\theta; \epsilon{\bf X}), \label{eq:gc_X} \\
G_{1}^{\mu} & = & \vb{\rho}_{0}\bdot \left( \mu\;\nabla\ln B \;+\; \frac{mv_{\|}^{2}}{B}\;\bhat\bdot\nabla\bhat \right) 
\;-\; \mu\;\frac{v_{\|}}{\Omega} \left( {\sf a}_{1}:\nabla\bhat \;+\; \bhat\bdot\nabla\btimes\bhat \right), 
\label{eq:gc_mu} \\
G_{1}^{\theta} & = & -\;\vb{\rho}_{0}\bdot{\bf R} \;+\; \pd{\vb{\rho}_{0}}{\theta}\bdot\nabla\ln B \;+\; 
\frac{v_{\|}}{\Omega}\;{\sf a}_{2}:\nabla\bhat \;+\; \frac{mv_{\|}^{2}}{2\,\mu B}\;\left( \bhat\bdot\nabla\bhat\bdot
\vb{\rho}_{0}\right), \label{eq:gc_theta}
\end{eqnarray}
and $G_{1}^{{\cal E}} \equiv 0$. Here, we use the rotating unit vectors $(\bhat,\wh{\bot},\wh{\rho})$: $\wh{\rho} = \bhat\btimes\wh{\bot} = \wh{{\sf 1}}\,\cos\theta - \wh{{\sf 2}}\,\sin\theta$, defined in terms of the fixed (local) unit vectors $\wh{{\sf 1}}\btimes\wh{{\sf 2}} = \bhat$, the vector field ${\bf R} = \nabla\wh{{\sf 1}}\bdot\wh{{\sf 2}}$ denotes Littlejohn's gyro-gauge vector field \cite{gc}, which is used to define the gradient operator
\begin{equation}
\nabla^{*} \;\equiv\; \nabla \;+\; \left[\; {\bf R} \;+\; \left( \frac{\bhat}{2}\bdot\nabla\btimes\bhat\right) \bhat \;\right] \pd{}{\theta}, 
\label{eq:grad_rho}
\end{equation}
and the gyroangle-dependent dyadic matrices are defined as 
${\sf a}_{1} = -\,\frac{1}{2} ( \wh{\rho}\,\wh{\bot} + \wh{\bot}\,\wh{\rho}) = \partial{\sf a}_{2}/\partial\theta$.

Using expressions for $G_{1}^{\alpha}$ and $G_{2}^{{\bf X}}$, the first-order correction $\vb{\rho}_{1}$ to the gyroradius vector $\vb{\rho}_{\epsilon}$ is given as 
\begin{equation}
\vb{\rho}_{1} = -\;\alpha\;\vb{\rho}_{0}\btimes\bhat - \beta\;\vb{\rho}_{0} - \gamma\;\bhat,
\label{eq:rho_1}
\end{equation}
where the coefficients $\alpha$, $\beta$, and $\gamma$ are defined as
\begin{eqnarray*}
\alpha & = & \frac{v_{\|}}{\Omega}\;{\sf a}_{2}:\nabla\bhat \;+\; \pd{\vb{\rho}_{0}}{\theta}\bdot 
\left( \nabla\ln B \;+\; \frac{mv_{\|}^{2}}{2\,\mu B}\;\bhat\bdot\nabla\bhat \right), \\
\beta & = & \vb{\rho}_{0}\bdot\left( \frac{1}{2}\;\nabla\ln B \;+\; \frac{mv_{\|}^{2}}{2\,\mu B}\;
\bhat\bdot\nabla\bhat \right) \;+\; \left. \left. \frac{v_{\|}}{2\Omega}\; \right( \bhat\bdot\nabla\btimes\bhat \;-\; 
{\sf a}_{1}:\nabla\bhat \right), \\
\gamma & = & 2\;\pd{\vb{\rho}_{0}}{\theta}\bdot \left( \frac{v_{\|}}{\Omega}\;\bhat\bdot\nabla
\bhat \right) \;-\; \frac{\mu B}{m\Omega^{2}} \left( \frac{1}{2}\,\bhat\bdot\nabla\ln B \;+\; {\sf a}_{2}:
\nabla\bhat \right).
\end{eqnarray*}
Using Eq.~(\ref{eq:rho_1}), the particle kinetic momentum (\ref{eq:p_epsilon}) can now be expressed as
\begin{eqnarray}
{\bf p}_{\epsilon} & = & {\bf p} \;+\; \epsilon \left[\; \frac{\bhat}{\Omega}\btimes\left( \mu\;\nabla B \;+\; mv_{\|}^{2}\,\bhat\bdot\nabla\bhat 
\right) \right. \nonumber \\
 &   &\mbox{}\hspace*{0.5in}-\; m\Omega \left(\pd{\gamma}{\theta} \;+\; \frac{v_{\|}}{\Omega}\;\bhat\bdot\nabla\bhat\bdot\vb{\rho}_{0} \right) \bhat \nonumber \\
 &   &\mbox{}\hspace*{0.5in}-\; m\Omega \left(\; \beta \;+\; \pd{\alpha}{\theta} \;-\; \frac{v_{\|}}{2\,\Omega}\;\bhat\bdot\nabla\btimes\bhat \;\right) 
\pd{\vb{\rho}_{0}}{\theta} \nonumber \\
 &   &\left.\hspace*{0.5in}+\; m\Omega\; \left(\; \alpha \;-\; \pd{\beta}{\theta} \;-\; \frac{v_{\|}}{2\,\Omega}\;\bhat\bdot\nabla\ln B \;\right) \vb{\rho}_{0} \;\right],
\label{eq:Peps_full}
\end{eqnarray}
where $\alpha$, $\beta$, and $\gamma$ are defined above. It is now a simple exercise using the formulas presented here to show that Eq.~(\ref{eq:Peps_full}) satisfies property (\ref{eq:Peps_prop}). Lastly, the gyroangle averages used in 
Sec.~IV.C [Eqs.~(\ref{eq:IFP_KDgc_mu})-(\ref{eq:IFP_Dgc_imu})] make use of the following expressions
\begin{eqnarray}
\left\langle {\bf p}_{\epsilon}\bdot\frac{\Omega}{\mu B}\;\pd{\vb{\rho}_{\epsilon}}{\theta} \right\rangle & = & 2 \;-\;
\frac{v_{\|}}{\Omega}\;\bhat\bdot\nabla\btimes\bhat, \label{muE_ifp} \\
\left\langle \left( {\bf p}_{\epsilon}\bdot\frac{\Omega}{\mu B}\;\pd{\vb{\rho}_{\epsilon}}{\theta}\right)^{2} \right\rangle & = & 4 \left( 1 \;-\; \frac{v_{\|}}{\Omega}\;\bhat\bdot\nabla\btimes\bhat\right), \label{mumu1_ifp} \\
\left\langle \left|\frac{\Omega}{B}\;\pd{\vb{\rho}_{\epsilon}}{\theta}\right|^{2}\right\rangle & = & \frac{2\mu}{mB}
\left( 1 \;-\; \frac{v_{\|}}{\Omega}\;\bhat\bdot\nabla\btimes\bhat\right), \label{mumu2_ifp} \\
\left\langle \left( {\bf p}_{\epsilon}\bdot\frac{\Omega}{\mu B}\;\pd{\vb{\rho}_{\epsilon}}{\theta}\right)
{\bf p}_{\epsilon} \right\rangle & = & 2\,p_{\|}\;\bhat \;+\; (m\,{\bf v}_{{\rm gc}})_{\bot}, \label{muspace_ifp}
\end{eqnarray}

\vfill\eject

\end{document}